\documentclass[
aip,
superscriptaddress,
amsmath,amssymb,
reprint,
floatfix
]{revtex4-1}

\usepackage{graphicx}% Include figure files
\usepackage{dcolumn}% Align table columns on decimal point
\usepackage{bm}% bold math
\usepackage[utf8]{inputenc}
\usepackage[T1]{fontenc}
\usepackage{mathptmx}
\usepackage{etoolbox}
\usepackage{gensymb}
\usepackage[colorlinks=true, linkcolor=blue, citecolor=blue, urlcolor=blue]{hyperref}% add hypertext capabilities
\usepackage[dvipsnames]{xcolor} %%https://www.overleaf.com/learn/latex/Using_colors_in_LaTeX

\makeatletter
\def\@email#1#2{%
 \endgroup
 \patchcmd{\titleblock@produce}
  {\frontmatter@RRAPformat}
  {\frontmatter@RRAPformat{\produce@RRAP{*#1\href{mailto:#2}{#2}}}\frontmatter@RRAPformat}
  {}{}
}%
\makeatother
\begin{document}

%\preprint{AIP/123-QED}

\title[Quantum anomalous Hall effect for metrology]{Quantum anomalous Hall effect for metrology}

\author{Nathaniel J. Huáng}
\altaffiliation{All authors contributed equally. Author to whom correspondence may be addressed: nathaniel.huang@npl.co.uk.}
\affiliation{Department of Quantum Technologies, National Physical Laboratory, Teddington TW11 0LW, United Kingdom}

\author{Jessica L. Boland}
\altaffiliation{jessica.boland@manchester.ac.uk}
\affiliation{Department of Quantum Technologies, National Physical Laboratory, Teddington TW11 0LW, United Kingdom}
\affiliation{Photon Science Institute and Department of Materials, University of Manchester, Manchester M13 9PL, United Kingdom}

\author{Kajetan M. Fijalkowski}
\altaffiliation{kajetan.fijalkowski@physik.uni-wuerzburg.de}
\affiliation{Institute for Topological Insulators and Faculty for Physics and Astronomy (EP3), Universität Würzburg, 97074 Würzburg, Germany}

\author{Charles Gould}
\altaffiliation{charles.gould@physik.uni-wuerzburg.de}
\affiliation{Institute for Topological Insulators and Faculty for Physics and Astronomy (EP3), Universität Würzburg, 97074 Würzburg, Germany}

\author{Thorsten Hesjedal}
\altaffiliation{thorsten.hesjedal@physics.ox.ac.uk}
\affiliation{Department of Physics, Clarendon Laboratory, University of Oxford, Oxford OX1 3PU, United Kingdom}

\author{Olga Kazakova}
\altaffiliation{olga.kazakova@npl.co.uk}
\affiliation{Department of Quantum Technologies, National Physical Laboratory, Teddington TW11 0LW, United Kingdom}
\affiliation{Department of Electrical and Electronic Engineering, University of Manchester, Manchester M13 9PL, United Kingdom}

\author{Susmit Kumar}
\altaffiliation{sku@justervesenet.no}
\affiliation{Justervesenet - Norwegian Metrology Service, 2007 Kjeller, Norway}

\author{Hansjörg Scherer}
\altaffiliation{hansjoerg.scherer@ptb.de}
\affiliation{Physikalisch-Technische Bundesanstalt, 38116 Braunschweig, Germany}

\date{\today}

\begin{abstract}
The quantum anomalous Hall effect (QAHE) in magnetic topological insulators offers great potential to revolutionize quantum electrical metrology by establishing primary resistance standards operating at zero external magnetic field and realizing a universal "quantum electrical metrology toolbox" that can perform quantum resistance, voltage and current metrology in a single instrument. To realize such promise, significant progress is still required to address materials and metrological challenges -- among which, one main challenge is to make the bulk of the topological insulator sufficiently insulating to improve the robustness of resistance quantization. In this \textit{Perspective}, we present an overview of the QAHE; discuss the aspects of topological material growth and characterization; and present a path towards a QAHE resistance standard realized in magnetically doped (Bi,Sb)$_2$Te$_3$ systems. We also present guidelines and methodologies for QAHE resistance metrology, its main limitations and challenges as well as modern strategies to overcome them.

\end{abstract}

\maketitle

\section{Introduction}
Electrical metrology is a domain that has been at the forefront of adopting new physics. Discovered more than a century ago \cite{hall}, the Hall effect underwent a transformative evolution when Klaus von Klitzing observed its quantized counterpart in 1980 in a two-dimensional electron gas system (2DES). This discovery of the quantum Hall effect (QHE) \cite{klitzing} linked the Hall resistance $R_H$ to universal SI constants \cite{Stock2019} -- Planck’s constant ($h$) and electron charge ($e$), with the von Klitzing constant $R_K$ defined as $R_H = \frac{R_K}{i} = \frac{h}{ie^2}$, where $i$ is an integer). This revolutionized resistance metrology by establishing the QHE as the primary resistance standard (PRS). Currently, high-quality 2DES to observe QHE are fabricated in GaAs/AlGaAs quantum well heterostructures grown by molecular beam epitaxy (MBE). Graphene, a two-dimensional single atomic layer of carbon, was quickly realized as promising PRS after its discovery \cite{Zhang2005, OK1}. Over the years, the metrology community has refined the production of epitaxial graphene and developed fabrication techniques suitable for resistance standards.  Graphene devices have shown remarkable agreement with the GaAs resistance standard maintained by the International Bureau of Weights and Measures, with a variance within 1~n$\Omega$/$\Omega$ and an uncertainty of $\pm$~3~n$\Omega$/$\Omega$ \cite{Chatterjee2023}.

The grand goal of electrical metrology is to realize a "quantum electrical metrology toolbox" where the three components of the quantum electrical metrology triangle -- the volt (V), ampere (A), and ohm ($\Omega$) -- exist in a single system. To achieve this, a "relaxed-condition" environment (including zero-to-low magnetic fields, higher operation temperatures and currents, and the ability to fabricate large-area devices \cite{Okazaki2022}) is critical for universal adoption of quantum electrical SI standards from metrological laboratories to industry. The quantum anomalous Hall effect (QAHE) intrinsically offers a path to circumvent the necessity to form Landau levels to realize a PRS with unsurpassed accuracy in such "relaxed conditions". This \textit{Perspective} presents recent advancement and current challenges for the realization of QAHE-based PRS, and provides strategies and methodologies towards a new era of quantum electrical metrology.

\section{Quantum anomalous Hall effect}
\subsection{Overview}
The idea of the QAHE dates back to 2003 and was first discussed in the context of 2D ferromagnets \cite{Onoda2003}. The breakthrough came with the advent of topological insulators (TI) \cite{Kane2005,Kane2005b,Bernevig2006,Moore2007,Fu2007,Fu2007b,konig2007,Hsieh2008,Hasan2010}, where the QAHE was first predicted to emerge in (Hg,Mn)Te quantum wells \cite{Liu2008}. There, the magnetic exchange interaction from the Mn atoms lifts the band inversion for one of the two spin species, effectively "removing" one of the two helical edge modes of the quantum spin Hall effect (QSHE) \cite{Kane2005,Kane2005b,Bernevig2006,konig2007,Konig2008}. A remaining single chiral edge mode is a realization of the QAHE with a Chern number $C = 1$. (Hg,Mn)Te is paramagnetic \cite{Furdyna1988}, which prevents it from realizing electronic quantization at zero external magnetic field \cite{Shamim2020}. Fortunately, another material family emerged: group V-VI ferromagnetic TIs (transition metal doped Bi$_2$Te$_3$, Sb$_2$Te$_3$, and Bi$_2$Se$_3$) \cite{Zhang2009,Yu2010}, followed by a long-awaited experimental realization of the zero external magnetic field QAHE in Cr-doped (Bi,Sb)$_2$Te$_3$ (BST) in 2013 \cite{Chang2013}, soon afterwards reproduced by multiple independent groups \cite{Checkelsky2014,Kou2014,Bestwick2015,Kandala2015,Chang2015b,Chang2015,Grauer2015}. Most notably, despite all the progress in the field, metrologically relevant precision of QAHE quantization has thus far been only reported in Cr/V-doped (Bi,Sb)$_2$Te$_3$ \cite{Goetz2018,Fox2018,Okazaki2020,Okazaki2022,Rodenbach2022,Patel2024}. A key point to appreciate is that the difficulty in achieving a pristine QAHE in a TI lies in controlling the bulk insulator rather than the topological surface state. The topological state is robust to perturbation. The bulk however is not protected and, given the relatively narrow band gaps which are ubiquitous in TIs, is very challenging to make sufficiently insulating. For a comprehensive understanding of the physical mechanisms, relevant material systems, and potential applications of the QAHE, the reader is referred to a recent review \cite{chang2023} by Chang \textit{et al}.

\subsection{Dimensionality and axion electrodynamics}
The original idea for realizing the QAHE from "removing" one of the two QSHE edge modes of a two-dimensional TI \cite{Liu2008} implies that the effect is inherently two-dimensional. Indeed, the conductivity tensor ($\sigma_{xx}$, $\sigma_{xy}$) scaling behavior analysis of the QAHE films \cite{Wang2014,Checkelsky2014,Kou2015,Kawamura2018} often reveal symmetries concurrent to that of conventional QHE observed in ordinary 2DESs \cite{Pruisken1985,Pruisken1988,Kivelson1992,Dykhne1994,Ruzin1995,Hilke1999}, without any non-trivial electrodynamics \cite{Fijalkowski2021a}. On the other hand, a perspective of the QAHE from the surface state of a three-dimensional TI film paints a more intriguing picture and involves axion electrodynamics \cite{Wilczek1987,Qi2008,Essin2009,Qi2011,Nomura2011,Wang2015,Morimoto2015}. Axionic corrections to Maxwell’s equations lead to a half-integer $\sigma_{xy} = \frac{e^2}{2h}$ quantization of Hall conductance from an individual topological surface \cite{Qi2008,Tse2010,Maciejko2010,Nomura2011,Wang2015,Morimoto2015}, and a signature of such half-integer quantization can be regarded as evidence for axion physics. This quantization has recently been extensively studied and debated in various electronic transport \cite{Mogi2017,Grauer2017,Mogi2018,Xiao2018,Wu2020,Fijalkowski2021a,Mogi2022,Zhuo2023b,Wang2024} and optical \cite{Wu2016,Okada2016,Dziom2017,Berger2022} experiments. Noteworthy, an apparent distinct QAHE origin in two- and three-dimensional magnetic TIs implies a distinct protection mechanism for a topological state in each, potentially carrying relevance for practical applications.

\subsection{Unusual magnetism}
The nature of interactions underpinning the magnetism remains an open question. Cr- and V-doped (Bi,Sb)$_2$Te$_3$ materials have similar Curie temperatures $\sim$20~K but vastly different strength of magnetic anisotropy \cite{Chang2013,Chang2015}. Various mechanisms were attempted to explain experimental observations, including Van Vleck ferromagnetism \cite{Yu2010,Kou2013,Chang2013,Chang2013b,Li2015,Wang2018}, Ruderman-Kittel-Kasuya-Yosida interactions \cite{Li2012,Kou2013,Wang2018}, and double-exchange/superexchange mechanisms \cite{Vergniory2014,Peixoto2016,Ye2019,Tcakaev2020}. While it appears that ferromagnetism plays an important role \cite{Chang2013,Chang2015,Wang2018}, there is evidence for superparamagnetic-like dynamics \cite{Lachman2015}, even concurrent with perfect electronic transport quantization near zero external magnetic field \cite{Grauer2015}. Electronic transport studies reveal rich magnetic domain driven phenomenology \cite{Qiu2022,Fijalkowski2022,Zhou2023} when the device size approaches a characteristic magnetic domain dimension \cite{Lachman2015,Wang2016,Yasuda2017,Wang2018}. Individual magnetic domains can have a ground state with magnetization pointing antiparallel relative to its neighbors \cite{Fijalkowski2022}, indicating a complex competition of ferro- and antiferromagnetic domain-domain interactions. In addition, other complex magnetic phenomena were reported on in this material system, including magnetic skyrmions \cite{Yasuda2016,Liu2017,Jiang2020}, coexistence of surface and bulk ferromagnetism \cite{Fijalkowski2020}, adiabatic cooling \cite{Bestwick2015}, sizable Barkhausen-like switching \cite{Liu2016}, and macroscopic quantum tunneling of the magnetization \cite{Fijalkowski2022}. It is not yet clear how all the experimental observations can be reconciled within a single unified model.

\subsection{Operational parameter space and experimental limitations}
The stability of QAHE in Cr/V-doped (Bi,Sb)$_2$Te$_3$ has been studied in detail. The explored parameter space includes the measurement current \cite{Kawamura2017,Fox2018,Rodenbach2021,Lippertz2022,Fijalkowski2024}, temperature \cite{Chang2013,Chang2015,Chang2015b,Li2016,Yasuda2020,Fijalkowski2021b}, material composition \cite{Mogi2015,Winnerlein2017,Ou2017}, and layer thickness \cite{Feng2016,Fijalkowski2021a,Zhuo2023b,Zhao2024}. The effect turns out to be surprisingly fragile, operating only in a narrow experimental window. Specifically to temperature, despite recent progress in increasing the operational temperature to $\sim$1~K \cite{Mogi2015,Ou2017,Yi2023}, a regime of metrologically relevant quantization remains limited to the mK range \cite{Goetz2018,Fox2018,Okazaki2020,Okazaki2022,Rodenbach2022,Patel2024}. The exact mechanism limiting the QAHE temperature is not yet fully understood, but it is likely to be related to formation of charge puddles \cite{Skinner2012,Chen2013,Nandi2018,Huang2021,Lippertz2022,Park2024} from material composition fluctuations, as the material is a mixed compound of \textit{n}-type Bi$_2$Te$_3$ and \textit{p}-type Sb$_2$Te$_3$ \cite{Zhang2011}. The concentration of magnetic doping can also fluctuate within the layer, which in addition to introducing magnetic inhomogeneities, can contribute to charge puddle formation if the magnetic dopants do not incorporate iso-electrically into the lattice. While the impact of these inhomogeneities can likely be reduced by further MBE development, the existence of charge puddles appears to be unavoidable due to inherently narrow bandgaps ubiquitous to TIs. Indeed, even state-of-the-art crystalline quality zinc-blende topological materials, such as HgTe, suffer from charge puddles affecting the electrical transport \cite{Thenapparambil2023,Fuchs2023}, despite more than 40 years of MBE growth development \cite{FAURIE1981} and 60 years of overall materials development \cite{LAWSON1959}. The position of the bulk valence band maximum in the band structure relative to the Fermi level could also play an important role \cite{Li2016}. In terms of the quality of QAHE quantization, the key parameter that needs to be optimized is the resistivity of the material between the edge channels \cite{Fijalkowski2021b}. The reports of edge state transport persisting up to the bulk Curie temperature \cite{Chang2015b,Yasuda2020,Fijalkowski2021b} imply that the topological state itself is robust, thus, further improvements are possible through materials optimization. For completeness we also mention that there has been some literature questioning the validity of a pure edge state description of the transport \cite{Rosen2022,ferguson2023}, but in the end, as long as the Hall resistance can be optimized close enough to perfect quantization, the effect can be used for metrology, regardless of the microscopic details of the current distribution within the device. On the issue of measurement current, the primary breakdown mechanism stems from the Hall electric field build-up when current is increased \cite{Kawamura2017,Fox2018,Rodenbach2021,Lippertz2022,Fijalkowski2024}. The build-up drives backscattering through the bulk and hinders the quantization. Recently, a new measurement scheme, "balanced quantum Hall resistor", aimed at eliminating this electric field has been demonstrated \cite{Fijalkowski2024,Patent}. Currently the metrology-grade experiments at zero external magnetic field remain limited to currents <100~nA \cite{Goetz2018,Fox2018,Okazaki2020,Rodenbach2022,Patel2024}. Application of an external magnetic field improves this to about 1~$\mu$A \cite{Okazaki2022}. Given that integer QHE based resistance standards routinely operate at currents >10~$\mu$A \cite{Ribeiropalau2015}, significant improvements of QAHE performance are necessary.

\subsection{Thin film growth}
\begin{figure}
\includegraphics[width=0.95\linewidth]{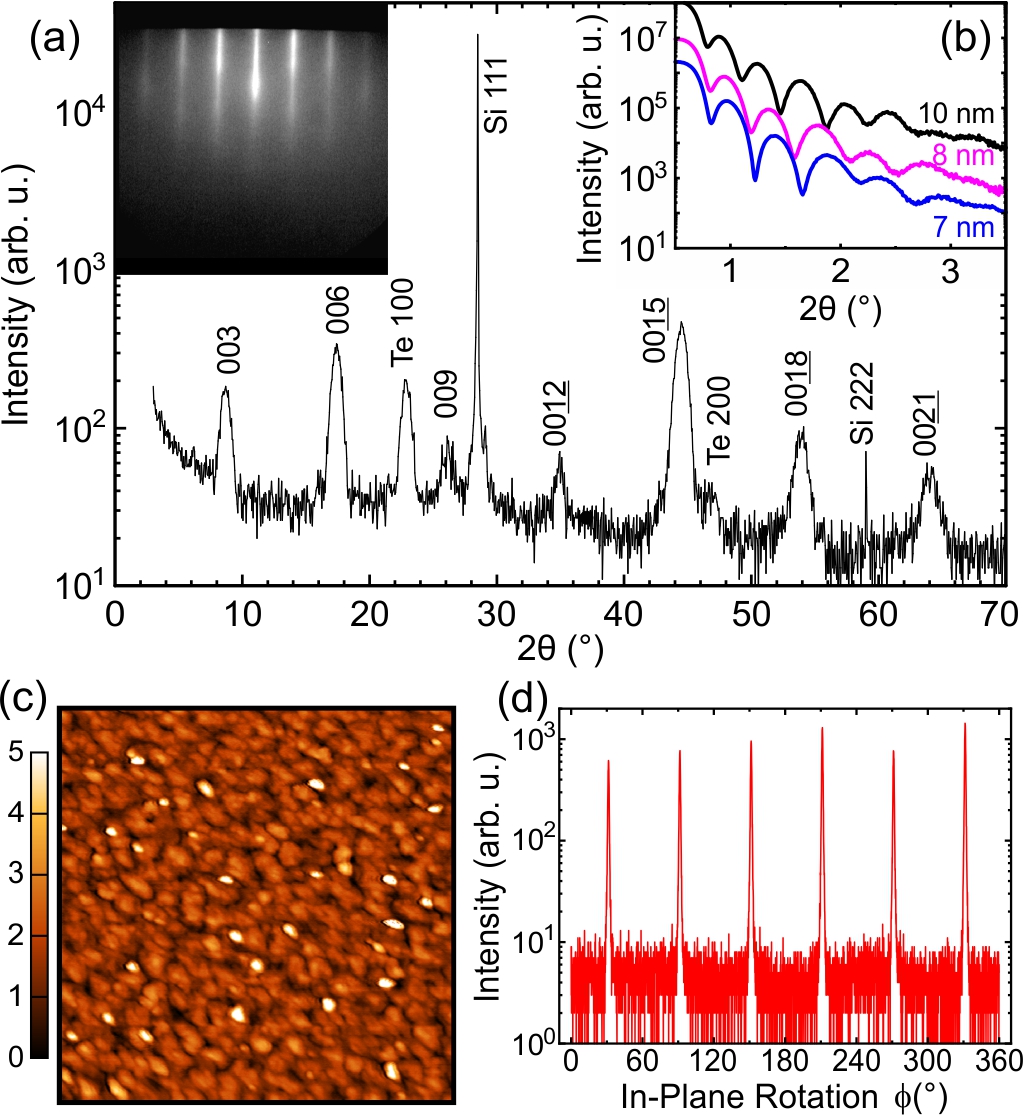}
\caption{\label{FIG1}
(a) XRD $\theta$-2$\theta$ scans of a V$_{0.1}$(Bi$_{0.21}$Sb$_{0.79}$)$_{1.9}$Te$_3$ film on Si (111). The film is 9~nm in thickness and capped with a 10-nm-thick Te layer to prevent the film from oxidation. The inset on the left depicts a RHEED pattern obtained from a similar layer's surface.
(b) XRR profiles for similar layers measuring 7~nm (blue), 8~nm (magenta), and 10~nm (black) in thickness, displayed with sequential offsets of a factor of 10 for visual clarity.
(c) $2 \times 2$ $\mu$m$^2$ AFM scan (5~nm z-scale) of an uncapped, 10-nm-thick film.
(d) $\phi$ scan of \{015\} reflections demonstrating the in-plane symmetry for a 9-nm-thick, Te-capped film on InP (111). Reprinted from Figures 1 and 2 with permission from Winnerlein \textit{et al}.\cite{Winnerlein2017} Copyright (2017) by the American Physical Society.}
\end{figure}

The Cr/V-doped (Bi,Sb)$_2$Te$_3$ thin films have been successfully grown using MBE on a wide variety of single-crystalline substrates, e.g., Al$_2$O$_3$ (0001) \cite{Taskin2012,Lee2012,Bansal2012,Harrison2013,Zhao2013}, SrTiO$_3$ (111) \cite{Chen_STO_2010, Zhang_STO_2011}, GaAs (111) \cite{Richardella2010,Chen2014,Eddrief2014,Liu2011}, Ge (111) \cite{Guillet2018, Kim2018}, Si (111) \cite{ZhangAPL2009,Li2010,He2011,Krumrain2011,Liu2013,Liu2012JVST} (Figure~\ref{FIG1}), CdS (0001) \cite{Kou2011}, and graphene \cite{Zhang2010NatPhys}. The lattice mismatch for TI films on these substrates is large, reaching $\sim$40\% for graphene \cite{He2013}. In fact, the governing van der Waals epitaxy growth mechanism does not require a perfect lattice match between film and substrate \cite{Koma1985,Koma1992}. The absence of strong ionic or covalent bonding across the film-substrate interface unavoidably results in some loss of control over the growth. The grains originating from randomly dispersed nucleation centers coalesce and form defects at their boundaries. For example, films on \textit{c}-plane sapphire are known to exhibit rotational twins, i.e., 6-fold symmetric reflection high-energy electron diffraction (RHEED) patterns, representative of the $R\bar{3}m$ crystal structure. In contrast, films grown on BaF$_2$ (111) are twin-free \cite{Bonell2017}. In general, the transport properties of TI films are most commonly not heavily dependent on the choice of substrate, but strongly affected by the generally large defect density at the interface \cite{Bansal2014}.

The deposition technique of choice for TI thin films is MBE. The growth of (Bi,Sb)$_2$Te$_3$ thin films by MBE is usually carried out with a considerable chalcogen overpressure on the order of 10:1 due to the high probability of re-evaporation \cite{Liu2021}, which can be lower if a chalcogen cracker cell is employed \cite{Zhang2012-cracker}. The growth rate is absorption controlled via the group-V element flux. The substrate temperature is the key parameter for controlling the film quality \cite{He2013}, which can be optimized further employing a two-step growth recipe, whereby a seed layer is first grown at a lower temperature \cite{Harrison2013}. Compensation doping has proven to be a successful strategy in electrically controlling defects. While Bi$_2$Te$_3$ can either be \textit{n}- or \textit{p}-type due to its ability to easily form both antisite defects and Te vacancies, Sb$_2$Te$_3$ is exclusively \textit{p}-type due to its inclination to form antisite defects. As a result, the synthesis of the compound (Bi$_{1-x}$Sb$_x$)$_2$Te$_3$ offers a direct approach to balance out \textit{n}- and \textit{p}-type conduction. When $x \approx 0.95$, the carrier density was lowered \cite{Zhang2011} to $2~\times~10^{12}$~cm$^{-2}$. This facilitated the observation of the QHE \cite{Yoshimi2015}, and ultimately the QAHE in Cr/V-doped BST \cite{Koma1985,Winnerlein2017,Yu2010,Chang2013,Checkelsky2014,Kou2014,Bestwick2015,Kandala2015,Chang2015b,Chang2015,Grauer2015}. Figure~\ref{FIG1} illustrates the structural properties of an optimally doped V$_{0.1}$(Bi$_{0.21}$Sb$_{0.79}$)$_{1.9}$Te$_3$ film on Si (111) \cite{Winnerlein2017}.

\section{Materials Characterization for QAHE resistance standard}
\begin{figure*}
\includegraphics[width=0.95\linewidth]{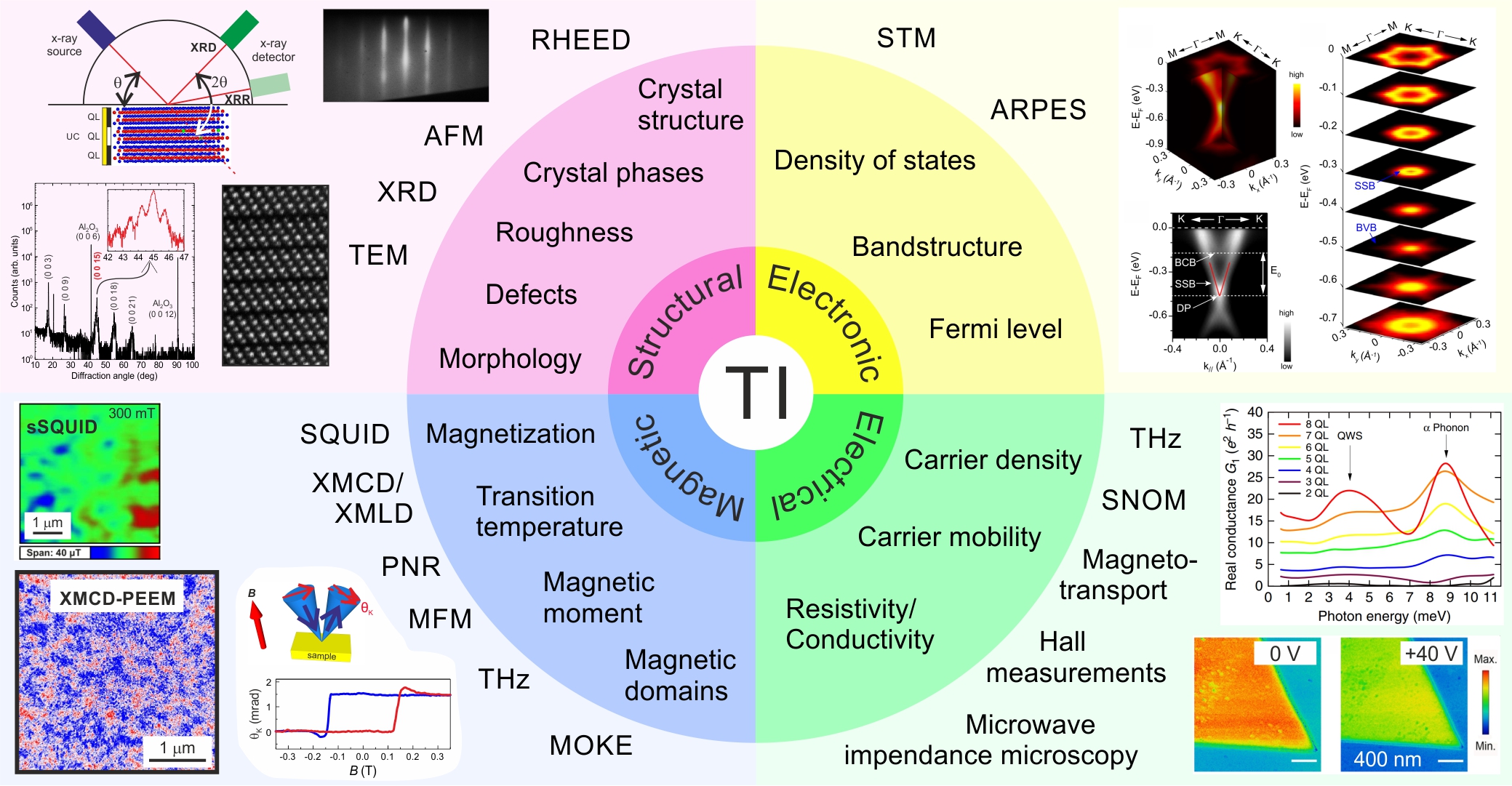}
\caption{\label{FIG2} Main classes of characterization techniques for QAHE materials, categorized into structural, electronic, electrical, and magnetic properties. Structural analysis relies on XRD, atomic force microscopy (AFM), transmission electron microscopy (TEM), and RHEED, each providing insights into crystal structures, phases, defects, and surface morphology. Electronic properties are explored using scanning tunneling microscopy (STM) and angle-resolved photoemission spectroscopy (ARPES), which examine the band structure, density of states, and the position of the Fermi level. Magnetic properties can be measured using techniques such as superconducting quantum interference device (SQUID) microscopy, X-ray magnetic circular dichroism photoemission electron microscopy (XMCD-PEEM), and magnetic force microscopy (MFM). Lastly, electrical properties can be assessed via scanning near-field optical microscopy (SNOM) and Hall measurements, providing carrier densities and mobilities. Together, these methods provide insight into the complex properties of TIs, essential for advancing their applications in technology.
Top (left and right) and bottom left insets are reprinted from Liu \textit{et al}. \cite{Liu2023} Bottom right insets are reprinted (adapted) with permission from Yuan \textit{et al}. \cite{yuan2017} Copyright 2017 American Chemical Society.}
\end{figure*}

To realize BST devices as QAHE PRS, deterministic engineering of electronic and magnetic properties of the surface and the bulk is essential. A feedback loop between growth and QAHE performance has to be established, including material characterization at both macroscopic and nanoscopic scales as illustrated in Figure~\ref{FIG2}. Inhomogeneities of structural, electronic, and magnetic properties at the nanoscale have been previously shown to have adverse effects on the robustness of QAHE quantization. Consequently, there is a need to combine bulk/macroscopic with surface-sensitive nanoscale characterization techniques, in order to distinguish bulk, surface, and edge properties, and to correlatively examine inhomogeneities that contribute to the high bulk conductivity.

\subsection{Structural characterization}
Structural properties of QAHE materials are commonly characterized \textit{in situ} using RHEED \cite{Harrison2014,Harrison2015massive} [inset to Figure~\ref{FIG1}(a)], as well as \textit{ex situ} techniques such as x-ray photoelectron spectroscopy (XPS) \cite{Ngabonziza2015,Peixoto2016,Leedahl2017}, x-ray diffraction and reflectometry (XRD, XRR) \cite{Reis2013,Collins-McIntyre2014,Haazen2012,Kashani2021} [Figure~\ref{FIG1}(a,b,d)], STM \cite{Liu2012}, and AFM \cite{Karma1998,Liu2023} [Figure~\ref{FIG1}(c)]. Combined with other \textit{ex situ} techniques, such as TEM with energy-dispersive X-ray spectroscopy (EDS) \cite{Harrison2015massive}, Rutherford backscattering spectroscopy (RBS) and particle-induced X-ray emission (PIXE) \cite{Harrison2015study}, as well as synchrotron-based extended X-ray absorption fine structure (EXAFS) \cite{Figueroa2015,Liu2014}, the stoichiometry, dopant states and concentration in TIs can be determined. Basic principles and instruments for these techniques have been comprehensively covered in a previous review \cite{Liu2023}, which demonstrated that structural properties can be identified from macroscopic down to atomic level.

\subsection{Electrical, electronic, and magnetic characterization}
Electrical and magnetic properties are typically only characterized \textit{ex situ}, which makes high lateral precision and especially tunability of the probing depth more demanding. Alternatively, development of \textit{in situ}, MBE-compatible electromagnetic characterization techniques would be advantageous, but remains a challenge. Given the challenges of reducing impurity concentrations and their effect on the electron potential landscape, it is important to measure the electronic band structure of TI thin films to determine the Fermi level and validate the existence of topological surface states. Electronic band structures of TI thin films are often measured by ARPES, which maps the energy- and angle-resolved distribution of valence band electrons emitted upon ionization via high-energy photon irradiation, enabling the band structure to be mapped along different crystallographic directions. As ARPES often uses incident light from a discrete gas resonance in the vacuum ultraviolet range \cite{Zhang2010NatPhys,Yang2018}, the probing depth is limited to the order of 1~nm. By tuning the incident photon energy utilizing synchrotron radiation or rapidly developing high-photon-energy laser sources, photon-energy-dependent measurements can be realized to extend the range of the probing depth to tens of nanometers, providing a means for distinguishing the surface and the bulk \cite{Hsieh2008,Xia2009,Chen2009,Chen2010}. However, the spatial resolution is limited to $\sim$50~$\mu$m in conventional ARPES, given by the typical incident beam size, and to hundreds of nanometers for nano-ARPES limited by the focus provided by a Fresnel zone plate \cite{Lim2024}, averaging over any nanoscale inhomogeneity.

At the macroscopic scale, low-precision magnetotransport can serve as an easily accessible pre-characterization technique to observe QAHE quantization and filter out devices with inferior performance. The selected devices can be further characterized in high-precision magnetotransport measurement systems operating at mK ranges currently required for metrology-grade QAHE quantization. Magnetotransport measurements can also provide spatially averaged values for inhomogeneities in electronic properties, such as characteristic quantum coherence lengths and energy scale of disorder potential due to charge puddles \cite{Alexander-Webber2018,Huang2015,Nandi2018}. While they can provide some insight into the effect of localized impurities, they are not surface-sensitive and cannot deliver spatially resolved analysis. Alongside magnetotransport, THz spectroscopy has emerged as a powerful non-contact tool for electrical characterization of TI thin films. The carrier scattering rates of TIs tend to fall within the THz frequency range, rendering THz radiation a low-energy probe capable of extracting carrier concentration, mobility, and lifetimes \cite{Joyce2016,Sim2014,Le2017,Ding2023,Kamboj2017,Sobota2012}. However, far-field THz spectroscopy is diffraction-limited and not surface-sensitive, therefore does not directly probe the TI surface and only infers spatially averaged properties aided via modeling.

Bulk magnetic properties of QAHE materials are commonly characterized using SQUID, vibrating-sample magnetometry, and magneto-optic Kerr effect \cite{Okada2016,De2021}. To be able to characterize local and depth-resolved magnetic properties, more advanced techniques such as XMCD and polarized neutron reflectometry are required. The recent review article \cite{Liu2023} provides a comprehensive overview of characterization techniques for magnetic properties of QAHE materials.

\subsection{Towards the nanoscale: scanning probe techniques}
Given that none of the techniques discussed above offers a tunable probing depth combined with nanoscale lateral resolution, we emphasize the importance of functional scanning probe microscopy (SPM), which allows local materials characterization at the nanoscale and can independently resolve edge, surface/subsurface, and bulk properties. Understanding nanoscale inhomogeneities of electronic and magnetic properties, including variations in charge carrier density, band bending effects, and the intricacies of magnetic structure, present significant challenges for QAHE metrology, making it a subject of considerable interest for the broader scientific community. There exists a multitude of complementary (opto-)electronic and magnetic SPM techniques, such as scattering-type SNOM \cite{Mooshammer2018,Pogna2021,Johnson2023}, scanning SQUID \cite{Romagnoli2023,Persky2022}, scanning microwave microscopy (SMM) \cite{Wang2023}, MFM, and Kelvin probe force microscopy (KPFM) \cite{Kim2021}. These methods intrinsically offer nanoscale spatial resolution typically below 50~nm, making them excellent choices for direct imaging of edge states and for nanoscale assessment of inhomogeneities in magnetically doped BST thin films and devices.

As an example, s-SNOM can perform optoelectronic characterization over a wide spectral range (visible-THz) beyond the diffraction limit \cite{Keilmann2004,Dai2018,Lloyd-Hughes2021,Chen2019} and its probing depth can be controlled via the nonlinear tip-sample interaction \cite{Taubner2005,Kusnetz2024,Moon2015,Mooshammer2018}. In conjunction with suitable modeling methods \cite{Keilmann2004,Cvitkovic2007,Hauer2012,McLeod2014,Mester2020,Vincent2024}, s-SNOM can in principle distinguish electronic states and conductivity with nanoscale (<20~nm) resolution both laterally and at different depths, hence useful for determining the effects of film thickness on hybridization of electronic states \cite{Mooshammer2018,Kastl2015,Ito2011,Mamyouda2015} and probing heterostructure interfaces and sub-surface dynamics. State-of-the-art s-SNOM can now operate at cryogenic temperatures below 10~K. Rapid instrumentational and metrological development \cite{Chen2019,Hu2022,Kim2023} provides a realistic prospect of operation at even lower temperatures and incorporation of high magnetic fields, yet there is further need for good practices of s-SNOM modeling, quantitative and traceable determination of key transport and electromagnetic parameters towards more robust materials metrology \cite{Leitenstorfer2023,Govyadinov2013,Chen2021}.

\section{QAHE Metrology}
\begin{figure}
\includegraphics[width=0.95\linewidth]{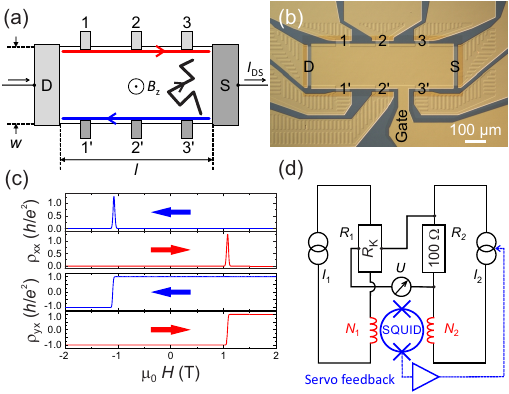}
\caption{\label{FIG3}
(a) Schematics of a QHRS device in Hall bar geometry (width $w$ and length $l$), with source (S), drain (D) and three pairs of Hall-voltage contacts (1-1' to 3-3'). The arrow represents the current flow for $I_{DS}$.
(b) Nomarski microscope image of a V$_{0.1}$(Bi$_{0.2}$Sb$_{0.8}$)$_{1.9}$Te$_3$ QAHE device \cite{Patel2024}. The gate electrode is used to tune the chemical potential in the TI film.
(c) Hall resistivity $\rho_{xy}$ and longitudinal resistivity $\rho_{xx}$ versus external magnetic field as measured for the V-doped BST Hall bar device. As typical fingerprint of the QAHE, the Hall resistivity is quantized in terms of $R_K$ for magnetic fields where the longitudinal resistivity vanishes (including at $B = 0$).
(d) Simplified schematic of a CCC resistance bridge used to measure the Hall resistance of a QAHE device.}
\end{figure}

The realization of a quantum Hall resistance standard (QHRS) based on conventional materials requires high magnetic fields generated by superconducting solenoids. This limits the operation practicality and makes it technically challenging to integrate with the Josephson voltage standards (JVS) in a single cryogenic system \cite{Brunpicard2016,Rodenbach2023}. QAHE in magnetically doped BST has demonstrated Hall resistance quantization without a permanent external magnetic field \cite{Yu2010,Chang2013,Chang2015,Bestwick2015,Chang2015b,Goetz2018,Fox2018,Okazaki2020,Rodenbach2022,Rodenbach2023,Patel2024}. Early transport measurements aiming at quantitative confirmation of Hall resistance quantization in the QAHE regime were performed on devices in Hall bar geometry [Figure~\ref{FIG3}(a,b)] using conventional equipment like calibrated resistors, voltage and current sources and detectors \cite{Bestwick2015,Chang2015b}. Since 2018, precision Hall resistance measurements at considerably higher accuracy levels were published, demonstrating Hall resistance quantization at the $10^{-6}$ level of relative uncertainty and better \cite{Goetz2018,Fox2018,Okazaki2020,Okazaki2022,Rodenbach2022,Rodenbach2023,Patel2024}.

\subsection{Cryogenic current comparators for resistance metrology}
The majority of these studies used metrology-grade high-precision resistance bridges based on the concept of the cryogenic current comparator (CCC) and included evaluations of the measurement uncertainties. CCC-based bridges and quantum-referenced standard resistors are well-established in resistance metrology and key to resistance measurements at state-of-the-art accuracy levels at ten parts-per-billion and below \cite{Poirier2009,Jeckelmann2001,Delahaye2003,K12}. As the goal is to measure a resistance ratio by comparing an unknown resistor with a resistance reference, the bridge techniques designed to measure ratios are inherently superior to potentiometric methods. A simplified measurement scheme with a CCC-based technique is shown in Figure~\ref{FIG3}(d). The CCC fixes the ratio of the currents $I_1$ and $I_2$, flowing through a QAHE device $R_1 = R_K$ and a calibrated reference resistor $R_2 = 100~\Omega$. The SQUID detects the magnetic flux proportional to the difference $N_1 \cdot I_1 - N_2 \cdot I_2$. The current $I_2$ is adjusted by a feedback loop controlled by the SQUID to null the difference, such that $I_1/I_2 = N_2/N_1$. The $R_1 / R_2$ ratio is calculated from the numbers-of-turns ratio $N_1/N_2$, from the bridge voltage difference $U$ and from the voltage drop $I_2 \cdot R_2$ across the reference resistor. Further details on advanced design and performance of state-of-the-art CCC bridges are given elsewhere \cite{Drung2009,Gotz2009,Drung2011,Drung2015}. Recent precision measurements on QAHE devices made from magnetically doped BST films were performed using bridges based on commercially available \cite{Goetz2018,Fox2018,Rodenbach2022,Rodenbach2023} or custom-made CCCs \cite{Okazaki2022,Patel2024,Patal2024a}. For Hall resistance measurements targeting highest accuracy and precision, the restriction to low bias currents ($\leq$100~nA) reflects in low fluxes generated by the windings of the CCC and coupling into the SQUID, which can limit the performance. Therefore, for measurements at very low currents the use of the highest possible numbers of turns $N_1$ and $N_2$ suitable for realizing $N_1/N_2 \approx R_2/R_1$ is favorable, resulting in minimizing the contribution of the SQUID to the uncertainty budget \cite{Patel2024}.

\begin{figure*}
\includegraphics[width=0.95\linewidth]{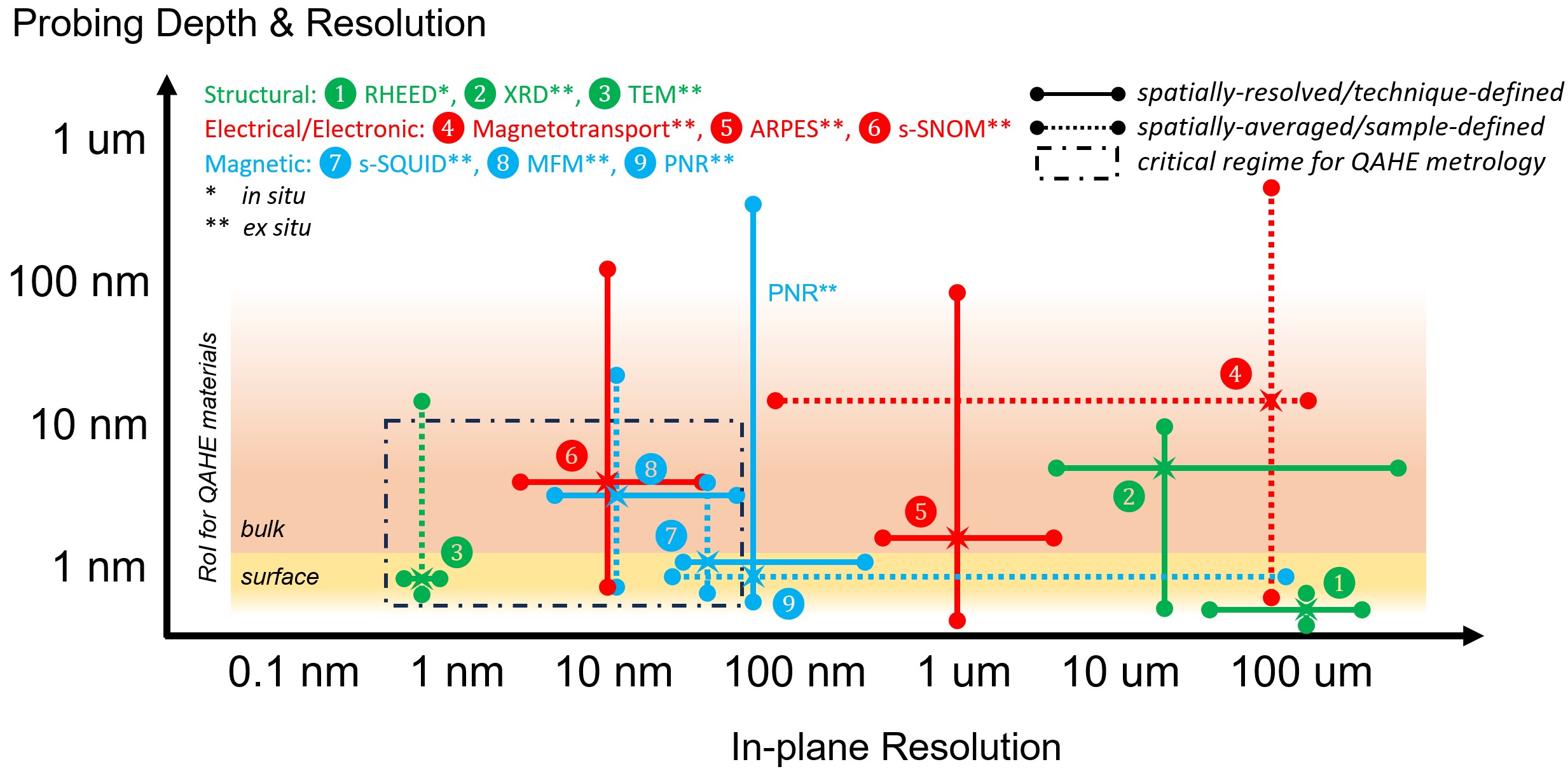}
\caption{\label{FIG4}
Comparative analysis of several essential characterization methods for QAHE materials with regards to probing depth and spatial resolution. Horizontal and vertical lines represent the tunability of in-plane and depth resolution, respectively; their intersections represent typical spatial resolution. The dash-dotted box highlights the critical regime required for the development of robust QAHE PRS.}
\end{figure*}

\subsection{Practical aspects: QHRS guidelines and methods}
General rules and best practices for the metrological application of the QHE in resistance metrology are laid out in technical guidelines \cite{Delahaye2003} for reliable dc measurements of the quantized Hall resistance. These guidelines describe the main tests and precautions necessary for both reproducible and accurate results in the use of the QHE to establish a reference standard of dc resistance with a relative uncertainty of a few parts in $10^9$. Although they are more specific for QHRS, the general methods and rules defined are also applicable to the QAHE resistance standard. QHRS devices are preferably shaped in a "Hall bar" geometry [Figure~\ref{FIG3}(a,b)]. In addition to the source and drain contacts for the bias current, they should be fitted with at least two pairs of contacts for measuring the Hall voltage, all of which providing galvanic connections to the conducting path of the device. For gated devices, an additional contact for applying a voltage to an electrostatic gate electrode is necessary, which should not have galvanic connection to the electron system.

Moreover, to have a reliable device, it is important that the Hall bar be large, such that the separation between opposite edges of the bar is much larger than the width of the edge state region. In this macroscopic limit, the main breakdown mechanism is scattering of electrons between conducting edge channels on different sides of the Hall bar, and therefore the critical current typically scales linearly with the width of the Hall bar, as observed in conventional QHRS devices \cite{Jeckelmann2001b}. For the QAHE in the macroscopic limit, it is believed that the breakdown mechanism is caused by electric-ﬁeld-driven percolation of charge puddles in the surface states of the TI films. Similar to conventional QHRS, a linear relationship between the critical current and the Hall bar width was observed \cite{Kawamura2017,Rodenbach2021,Lippertz2022}. Thus, the width of the Hall bar should be sufficiently large (100-800 $\mu$m), with the maximum size usually limited by the homogeneity of the TI films \cite{Goetz2018,Fox2018,Okazaki2020,Okazaki2022,Rodenbach2022,Rodenbach2023,Patel2024}. The source and drain current contacts should extend over the whole width of the device to maximize the critical current. Narrow voltage contacts should be avoided, as this can lead to non-equilibrium populations of the electronic edge states and partial depletion of the 2DES in the narrow channels, which will cause deviations of the Hall voltage from the nominal quantized value $R_K$ in the QAHE regime. The distances between the voltage contact pairs should be sufficiently large, i.e., in the order of the width of the device \cite{Delahaye2003}.

Another important aspect is quality of the device fabrication. While initially the experimental reports of QAHE often relied on mechanical scratching to define Hall bar-shaped structures \cite{Chang2013}, recent metrology-grade experiments rely solely on lithographically patterned devices \cite{Goetz2018,Fox2018,Okazaki2020,Okazaki2022,Rodenbach2022,Patel2024}. A top gate is employed to electrostatically tune the Fermi level and minimize the parasitic bulk conductivity, enabling pure edge state transport (a back gate should in principle work equally well for this purpose, but that has not yet been tested to metrological precision). While details of gate geometry are not expected to be relevant for DC resistance metrology, at AC frequencies, circuit capacitance considerations will obviously have to include the exact gating geometry.

The key feature of QHEs (including QAHE) is the occurrence of Hall (transverse) resistance quantization together with vanishing longitudinal resistance, when entering the regime of dissipationless transport along the edge channels [Figure~\ref{FIG3}(c)]. Realizing this transport regime requires adjusting the chemical potential so that bulk conductivity is suppressed to the minimum. The respective tuning of QHRS devices is preferably carried out by minimizing the longitudinal resistance (or voltage) measured along the device [on two opposite pairs of voltage contacts, 1-3 or 1'-3' in Figure~\ref{FIG3}(a,b)]. More accurate results are obtained by carrying out measurements of transverse voltages on several contact pairs including "orthogonal" and "diagonal" contacts, and then by calculating the longitudinal voltage by using Kirchhoff’s voltage law \cite{Delahaye2003}.

Non-ideal voltage contacts to the electron system can severely affect precision measurements and are critical for the device performance in metrology \cite{Delahaye2003}. Besides causing excessive current or voltage noise, poor voltage contacts can generate offset voltages which introduce systematic errors in measurements of the Hall resistance due to noise rectiﬁcation processes. As poor contacts are typically indicated by elevated contact resistance (and, in the worst case, by non-linear current-voltage characteristics), contact resistances should be determined prior to precision measurements of the QAHE. A practical method to perform contact resistance measurements uses a three-terminal measurement technique \cite{Delahaye2003}. For conventional QHRS devices, typical contact resistances of high-quality contacts are of the order 10~$\Omega$ or below; however, higher contact resistances up to 100~$\Omega$ are acceptable \cite{Delahaye2003}. For magnetically doped BST QAHE devices, contact resistances as low as several ohm were reported \cite{Fox2018,Okazaki2022,Patel2024,Patal2024a}. For achieving low contact resistances, choosing a split-finger design of the contacts implementing a miniature on-chip multiple terminal connection can be useful \cite{Delahaye1993}, as demonstrated for graphene-based devices \cite{Kruskopf2019}.

Besides the determination of contact resistances, the quantitative evaluation of the residual longitudinal resistivity $\rho_{xx}$ is key to the conditions of Hall resistance quantization in all known types of QHE devices. Ideally, the value of the transverse resistivity $\rho_{xy}$ on the resistance plateau is, within the limit of the resolution of the measurements, invariant over appreciable ranges of relevant experimental parameters like bias current or temperature. If bias current or temperature are increased beyond critical limits, the breakdown of the QHE with a deviation of $\rho_{xy}$ from the quantized value occurs \cite{Jeckelmann2001,Delahaye2003}. Generally, this is indicated by a gradual increase of $\rho_{xx}$. The dependence of $\rho_{xy}$ on $\rho_{xx}$ typically depends on the device geometry, the operating conditions, the magnetic ﬁeld direction, the position on the resistance plateau, and the bias current \cite{Poirier2009,Jeckelmann2001,Delahaye2003}. In QHRS devices, it has often been observed that $\rho_{xy}$ varies linearly with $\rho_{xx}$ over several decades in $\rho_{xx}$, described by the relationship $\delta\rho_{xx} = s \cdot \rho_{xx}$, with $\delta\rho_{xy} = \rho_{xy}/R_K - 1$ being the deviation of the Hall resistivity $\rho_{xy}$ relative to $R_K$ \cite{Cage1984,Diorio1986,Delahaye1986,Yoshihiro1986,Furlan1998,Jeckelmann2001,Delahaye2003}. In case of thermal activation, this behavior stems from the temperature dependence of $\rho_{xx}$ and the geometrical mixing of the longitudinal voltage into the Hall voltage. In conventional QHRS devices, experimentally determined values of the proportionality factor $s$ typically vary between -1 and -0.01. For QAHE devices made from magnetically doped BST films, values between about -0.1 and -0.01 were reported \cite{Okazaki2022,Patel2024}. In some cases, deviations from a linear $\rho_{xy}$($\rho_{xx}$) relationship were observed \cite{Fox2018,Patal2024a}. The dependence of $\rho_{xy}$ on $\rho_{xx}$ is of particular significance in quantum Hall resistance metrology. Analysis of $\delta\rho_{xy}$($\rho_{xx})$ when extrapolated to $\rho_{xx} = 0$ (dissipationless transport) \cite{Cage1984,Delahaye1986,Hartland1992,Furlan1998,Jeckelmann2001,Schopfer2013,Lafont2015} yields a proper measure for Hall resistance quantization under ideal conditions, allowing QAHE device performance evaluation for QHRS applications \cite{Fox2018,Okazaki2022,Patel2024}.

\section{Metrological outlook for QAHE}
\subsection{Materials optimization and metrological characterization}
The necessity of introducing magnetic order in TI films has spurred extensive research aimed at identifying the optimal dopant -- one that introduces magnetic order, while not being detrimental to the electronic properties. Both transition metal (TM) and rare earth doped TI films exhibit promising characteristics \cite{Hesjedal2019}. While ferromagnetic order has been reported in a number of TM-doped TIs, Cr and V possess the most robust long-range ferromagnetic order with the desired out-of-plane anisotropy and a typical $T_C$ of 59~K (104~K) for Cr (V) concentrations of $x =$ 0.29 (0.25) in Sb$_{2-x}$(V/Cr)$_x$Te$_3$ \cite{Chang2015}. As the QAHE is limited to temperatures far below their ferromagnetic ordering temperatures, magnetic doping is still an open material challenge.

To address the challenge of the non-insulating bulk, Figure~\ref{FIG4} compares several essential characterization methods required for QAHE materials. Their variations in probing depth, spatial resolution, and their alignment with the critical regime of interest for robust QAHE resistance standards are shown. Most current techniques do not offer a tunable probing depth on the nanometer regime -- a central obstacle for QAHE materials optimization. While s-SNOM represents a prime example of a tomographic technique with nanoscale resolution and tunability in all three dimensions, its application is confined in electronic and optical characterization. Therefore, we emphasize the importance of integrating complementary, multifunctional, and multiscale techniques. Concurrently, there is a need for development of \textit{in-situ}, local and depth-sensitive electronic and magnetic characterization methods, which can also reduce material degradation and contamination that can occur during \textit{ex situ} material evaluations. To further develop and reliably employ these techniques, while ensuring high accuracy and precision, metrological improvements to most of the abovementioned techniques are needed, namely: establishing practical standards of measurements, modeling techniques, and reference materials; defining and assuring unbroken chains of traceability to practical realizations of the SI units; better understanding the sources of variability and inaccuracy, improving reproducibility and comparability. We emphasize that only with these rigorous metrological approaches, we can progress metrological materials characterization to enable real-world innovations of TIs with confidence.

\subsection{Vision, strategies, and challenges for practical QAHE metrology}
Main motivations for the development of a QHRS operatable at zero external magnetic field are, \textit{i}) it is important for practical quantum electrical metrology, providing the attractive possibility to integrate a QHRS and a JVS in a single cryogenic apparatus, without hampering the operation of the JVS by magnetic fields needed for the QHRS operation; \textit{ii}) it is of an economical nature, i.e., to save the costs and technical efforts associated with the operation of high-field magnets, enabling wider dissemination of electrical units directly to end users and incorporation of precision metrology into routine calibrations in different economic sectors. The full practical and economic advantages of such integrated systems will be exploited using cryogen-free systems, such as a pulse-tube cryocooler with closed helium or helium-free circuits and small laboratory footprint. Commercial state-of-the-art cryocoolers can achieve temperatures down to about 2~K in combination with magnets sufficient to magnetize the QAHE device prior to operation. Such compact system would realize a universal "quantum electrical metrology toolbox" and offer the possibility to perform resistance, voltage and current metrology based on the primary quantum electrical effects. This strategic target is at the heart of modern quantum electrical metrology, which aims to develop and provide user-friendly, economic quantum standard systems for an extended usership suitable for direct exploitation by industry. Presently, several metrological institutes worldwide are working toward this direction. The National Institute of Standards and Technology (USA) recently implemented a quantum current sensor setup that combines a QAHE device and a JVS in a single cryostat \cite{Rodenbach2023}. With this, currents within the range 10-250~nA were measured with relative uncertainty 4-40 parts per million. However, the operation of QAHE device yet requires a dilution refrigerator, therefore, currently does not meet the desired requirements for practicality, affordability, or mobility. Furthermore, Rodenbach \textit{et al}. \cite{Rodenbach2023} reported significant heating of the quantum anomalous Hall device due to microwave radiation leakage from the programmable Josephson voltage standard in their setup, demonstrating that complications can arise from integrating a QAHE device with a JVS in a single dilution refrigerator. Generally, such integration is challenging since dilution refrigerators with millikelvin base temperatures have cooling powers limited to a few hundreds of microwatts typically, which imposes technical constraints on the acceptable heat load introduced by the wiring or by radiation sources of the system. These limitations are less severe, for instance, for pulse-tube cryocooler systems which offer cooling powers of the order of a watt typically. However, due to the temperature limitations set by state-of-the-art QAHE devices, the integration with the JVS in a cryocooler is currently not possible.

Regarding one of the main metrological benchmarks  -- the quantization accuracy -- state-of-the-art QAHE devices have recently demonstrated performance at the $10^{-9}$ level in measurements performed at the National Metrology Institute of Japan and at the Physikalisch-Technische Bundesanstalt (Germany) \cite{Okazaki2022,Patel2024}. This complies with the needs of metrological applications \cite{Poirier2009,Jeckelmann2001,Delahaye2003,K12} and holds promise for future further developments. Regarding the matter of increasing the applicable bias currents to the metrologically desired level beyond few $\mu$A \cite{Okazaki2022}, novel concepts based on advanced device layouts and operation schemes were proposed, and proof-of-principle experiments already showed promising results \cite{Patent,Fijalkowski2024}. Currently, these schemes are undergoing validation at accuracy levels relevant for metrological applications.

Most pressing need for practical QAHE metrology and resistance standards, therefore, is the remaining task to overcome the limitations regarding temperatures requiring dilution refrigerator operation. The current consensus is that this will also require significant improvements in the field of materials science and technology to explore novel material systems beyond the family of magnetically doped BST. These routes toward QAHE realizations at temperatures of at least 1~K are consequently pursued as part of the current European Partnership on Metrology project "Quantum Anomalous Hall Effect Materials and Devices for Metrology" \cite{JRP}.

\begin{acknowledgments}
The authors would like to acknowledge the support via the project "Quantum Anomalous Hall Effect Materials and Devices for Metrology" (QuAHMET). The project (23FUN07 QuAHMET) has received funding from the European Partnership on Metrology, co-financed from the European Union’s Horizon Europe Research and Innovation Programme and by the Participating States. N.J.H., J.L.B., and O.K. acknowledge the support of the UK government Department for Science, Innovation and Technology through the UK National Quantum Technologies Programme. J.L.B also acknowledges UKRI the financial support via her Future Leaders Fellowship (MR/T022140/1); EPSRC via the NAME program grant (EP/V001914/1) and EP/T01914X/1; and the Henry Royce Institute for their support. H.S. gratefully acknowledges the financial support by the European Commission under the H2020 FETPROACT Grant TOCHA (824140), by EMPIR 20FUN03 COMET (this project has received funding from the EMPIR program co-financed by the Participating States and from the European Union’s Horizon 2020 research and innovation program), and by the Deutsche Forschungsgemeinschaft (EXC-2123 QuantumFrontiers, 390837967). K.M.F. and C.G. gratefully acknowledge the financial support of the Free State of Bavaria (the Institute for Topological Insulators), Deutsche Forschungsgemeinschaft (SFB 1170, 258499086), Würzburg-Dresden Cluster of Excellence on Complexity and Topology in Quantum Matter (EXC 2147, 39085490), and the European Commission under the H2020 FETPROACT Grant TOCHA (824140).

\end{acknowledgments}

\section*{Conflict of Interest}
The authors have no conflicts to disclose.

\section*{Data Availability}
The data that support the discussions of this article are available from the corresponding authors upon reasonable request.

\section*{Authors Contributions}
\textbf{N.J.H.}: Conceptualization (equal); Data Curation (equal); Formal Analysis (equal); Funding Acquisition (equal); Investigation (equal); Methodology (equal); Resources (equal); Visualization (equal); Writing/Original Draft Preparation (equal); Writing/Review and Editing (equal); Project Administration (lead). \textbf{J.L.B.}: Conceptualization (equal); Data Curation (equal); Formal Analysis (equal); Funding Acquisition (equal); Investigation (equal); Methodology (equal); Resources (equal); Visualization (equal); Writing/Original Draft Preparation (equal); Writing/Review and Editing (equal). \textbf{K.M.F.}: Conceptualization (equal); Data Curation (equal); Formal Analysis (equal); Investigation (equal); Methodology (equal); Resources (equal); Visualization (equal); Writing/Original Draft Preparation (equal); Writing/Review and Editing (equal). \textbf{C.G.}: Conceptualization (equal); Data Curation (equal); Formal Analysis (equal); Funding Acquisition (equal); Investigation (equal); Methodology (equal); Resources (equal); Visualization (equal); Writing/Original Draft Preparation (equal); Writing/Review and Editing (equal). \textbf{T.H.}: Conceptualization (equal); Data Curation (equal); Formal Analysis (equal); Funding Acquisition (equal); Investigation (equal); Methodology (equal); Resources (equal); Visualization (equal); Writing/Original Draft Preparation (equal); Writing/Review and Editing (equal). \textbf{O.K.}: Conceptualization (equal); Data Curation (equal); Formal Analysis (equal); Funding Acquisition (equal); Investigation (equal); Methodology (equal); Resources (equal); Visualization (equal); Writing/Original Draft Preparation (equal); Writing/Review and Editing (equal). \textbf{S.K.}: Conceptualization (equal); Data Curation (equal); Formal Analysis (equal); Funding Acquisition (equal); Investigation (equal); Methodology (equal); Resources (equal); Visualization (equal); Writing/Original Draft Preparation (equal); Writing/Review and Editing (equal). \textbf{H.S.}: Conceptualization (equal); Data Curation (equal); Formal Analysis (equal); Funding Acquisition (equal); Investigation (equal); Methodology (equal); Resources (equal); Visualization (equal); Writing/Original Draft Preparation (equal); Writing/Review and Editing (equal).

%\nocite{*}
\bibliography{main.bib}

\end{document}